\documentclass[aps,prl,twocolumn,superscriptaddress,longbibliography,footinbib,nobibnotes,10pt,floatfix]{revtex4-2}  
\usepackage{preamble}
\newcommand{\prlsection}[1]{\textit{#1}---}
\begin{document}
\title{Long-Range Order and Composite Boson Condensation in Lattice Quantum Hall States}
\author{F.~J.~Pauw\orcidlink{0009-0006-4188-8503}}
\email{fabian.pauw@lmu.de}
\affiliation{\ascaddress}
\affiliation{\mcqstaddress}
\author{N.~Goldman}
\affiliation{\lkbaddress}
\affiliation{\ulbaddress}
\affiliation{\solvayaddress}
\author{F.~A.~Palm\orcidlink{0000-0001-5774-5546}}
\email{felix.palm@ulb.be}
\affiliation{\ulbaddress}
\affiliation{\solvayaddress}

\date{\today}

\begin{abstract}
    	Topological states of matter, such as quantum Hall states, are characterized by the absence of local order parameters.
    	In the continuum, their topological order has been reinterpreted as the condensation of nonlocal composite bosons, but this condensation and the resulting long-range correlations have so far eluded direct confirmation beyond the simplest trial states.
        Here, we apply tensor-network methods to lattice Hamiltonians hosting quantum Hall states, revealing this exotic order directly in interacting ground states and demonstrating both the long-range correlations and the condensation of composite bosons.
		These correlations define a nonlocal order parameter that identifies phase transitions between trivial and topological phases, for both bosonic and fermionic models.
    	The corresponding operators are accessible through site-resolved, local measurements on an extensive number of particles, making them directly probable in quantum simulators.
    	Our work opens new avenues for microscopic studies of topological quantum matter beyond the reach of traditional solid-state approaches.
\end{abstract}

\maketitle

\prlsection{Introduction}
Local order parameters and broken symmetries have long been cornerstones of condensed-matter physics, in particular to characterize phases of matter~\cite{Landau1937,Ginsburg1950}.
Common lore assumes that topologically ordered phases are exempt from this paradigm due to their lack of long-range order of local operators.
Nevertheless, generalizing order parameters from strictly local to nonlocal constructions allowed deeper insight into certain phases of topological quantum matter via, among others, string order parameters~\cite{denNijs1989,Kennedy1992,DallaTorre2006} and Wegner-Wilson loop operators~\cite{Wegner1971,Wilson1974,Fredenhagen1983}.

More recently, the framework of higher-form symmetries~\cite{Gaiotto2015,McGreevy2023,Pace2023} revealed that these nonlocal operators can be understood as the fundamental charged objects of an emergent topological symmetry, effectively extending the Landau paradigm of symmetry breaking to topological order.
In the context of fractional quantum Hall (FQH) states~\cite{Tsui1982}, early work by Girvin and MacDonald~\cite{Girvin1987,Girvin1988} reinterpreted analytical trial states in terms of composite bosons (CBs) -- bound states of a particle and flux quanta -- which exhibit off-diagonal quasi-long-range order and quasi-condensation.
Shortly after, Read~\cite{Read1989} noted that attaching quasiholes (vortices) instead of pure flux establishes genuine long-range order.
Later on, these concepts were extended using purely second-quantized formulations of FQH trial states~\cite{Mazaheri2015,Chen2019,Zhang2023}.
In all cases, spatially-resolved information about an extensive number of particles is required, making such approaches intractable in solid-state experiments in the thermodynamic limit.

In contrast, recent advances in quantum simulation of FQH systems with cold atoms in optical lattices~\cite{Leonard2023} and photons in superconducting circuits~\cite{Wang2024} have opened new opportunities of site-resolved studies of few-particle FQH states on lattices.
While such systems neither realize continuum FQH states nor their parent Hamiltonians, previous numerical works showed robustness of the quasi-long-range order of Girvin~and~MacDonald's CBs in finite-size lattices with sizable band dispersion~\cite{Pauw2024,Pauw2026}.
Despite this progress, fundamental gaps remain.
To date, lattice studies have exclusively focused on CBs of constituent particles and flux quanta.
While these exhibit algebraically decaying correlations, they fail to capture the genuine long-range order anticipated by Read~\cite{Read1989}.
Consequently, it has remained unclear whether such long-range order in FQH states can be resolved using site-resolved microscopy in quantum simulators.
Furthermore, direct evidence of CB (quasi-)condensation beyond trial states has been lacking for both types of composita, obfuscating the physical reality of the underlying mechanism.

\begin{figure}[t]
    \centering
    \includegraphics{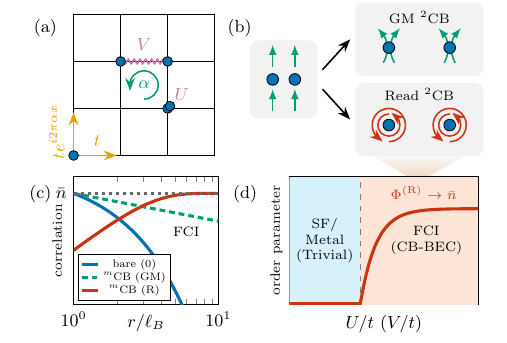}
    \caption{
    (a) Sketch of the square-lattice interacting Hofstadter model with on-site (Hubbard) repulsion $U$ (bosons) or nearest-neighbor repulsion $V$ (fermions).
    (b) In the topological phase at filling factor $\nu\!=\!\nicefrac{1}{m}$ (here, $m\!=\!2$), both Read's quasihole mapping (upper panel) and Girvin~and~MacDonald's (GM) flux attachment (lower panel) yield composite bosons ($^m$CBs).
    (c) While the correlations of the bare particles decay rapidly with distance $r$ (blue), GM's CBs exhibit algebraic quasi long-range order (green), while Read's CBs exhibit true long-range order (red).
	(d) In the bulk, we define a nonlocal order parameter $\Phi^{(R)}$, resolving phase transitions between trivial and topological phases.
    }
    \label{fig:Sketch}
\end{figure}

In this Letter, we present a unified composite lattice boson approach that establishes genuine long-range order in lattice quantum Hall states.
By explicitly confirming the (quasi-)condensation of CBs, we define a nonlocal order parameter and apply it to pinpoint phase transitions between topologically trivial and topologically ordered states in the interacting Harper-Hofstadter model.
Crucially, our lattice approach promotes these nonlocal observables from merely theoretical constructs to experimentally accessible signatures, paving the way for the detection of topological order via CB condensation.
\prlsection{Model and methods}
A paradigmatic lattice model of the quantum Hall problem is the Harper-Hofstadter model with single-particle Hamiltonian
\begin{equation}\label{eq:model}
    \hcH = -t\sum_{\langle \vec{j}\vec{k} \rangle} \left(\de^{\di \phi_{\vec{j}\rightarrow \vec{k}}} \had_{\vec{j}} \ha_{\vec{k}} + \mathrm{h.c.}\right),
\end{equation}
where $\langle ...\rangle$ denotes nearest neighbors, $\hat{a}_{\vec{j}}^{\dagger}$ ($\hat{a}_{\vec{j}}^{\nodagger}$) creates (annihilates) a spinless boson or fermion at the site \mbox{$\vec{j}\!=\!(j_x,j_y)$}, and $\hat{n}_{\vec{j}}\!=\!\had_{\vec{j}}\ha_{\vec{j}}$ is the particle number operator.
Here, we consider this model on an $L \times L$ square lattice with open boundary conditions and lattice constant $a$ [\cref{fig:Sketch}(a)].
The complex phase implements a uniform magnetic flux background $2\pi \alpha \!=\! \sum_{\square} \phi_{\vec{j}\rightarrow \vec{k}}$ per plaquette.
Such a magnetic flux introduces the magnetic length $\ell_B \!=\! \nicefrac{1}{\sqrt{2\pi\alpha}}$ as a characteristic length scale competing with the lattice length.
In the low-flux limit, $\alpha\!\to\!0$, one recovers the continuum Landau level regime.
It has been well established that this model and extensions thereof host a variety of lattice quantum Hall (QH) states.
Apart from non-interacting Chern insulators (CI) for fermions at integer filling factors, interactions can stabilize fractional Chern insulators (FCIs) akin to the Laughlin states~\cite{Laughlin1983} at filling factor $\nu\!=\!\nicefrac{\bar{n}}{\alpha}\!=\!\nicefrac{1}{m}$ for both bosons ($m$ even) and fermions ($m$ odd), where $\bar{n}$ is the particle number density in the uniform bulk.
For instance, on-site Hubbard repulsion,
\begin{equation}\label{eq:boson_model}
    \hcH_{\rm B} =  \frac{U}{2} \sum_{\vec{j}}\hat{n}_{\vec{j}}(\hat{n}_{\vec{j}}-1),
\end{equation}
stabilizes a bosonic Laughlin state at filling fraction \mbox{$\nu\!=\!\nicefrac{1}{2}$} for sufficiently strong on-site repulsion, \mbox{$U/t\!>\!0$}~\cite{Soerensen2005,Hafezi2007,Gerster2014,Boesl2022,Wang2022,Palm2022,Leonard2023,Wang2024}.
For the fermionic case, nearest-neighbor repulsion, $V/t \!>\! 0$,
\begin{equation}\label{eq:fermion_model}
    \hat{\mathcal{H}}_{\mathrm{F}} = V \sum_{\langle \vec{j} \vec{k} \rangle }\hat{n}_{\vec{j}}\hat{n}_{\vec{k}},
\end{equation}
stabilizes the $\nu\!=\!\nicefrac{1}{3}$-Laughlin state~\cite{Motruk2017,Pauw2026}.

To treat all systems on equal footing, we employ density matrix renormalization group (DMRG) methods to variationally find a matrix product state (MPS) representation of the ground state of the model~\cite{White1992,Schollwock2005,Schollwock2011}.
We exploit particle number conservation and systematically increase the bond dimension up to $\chi\!=\!2048$, at which our observables are converged~\cite{SuppMat}.

\prlsection{Composite lattice bosons}
A hallmark of topologically ordered phases is the rapid decay of the single-particle correlations,
\begin{equation}\label{eq:bare_correlator}
    \rho^{(0)}_{\vec{j},\vec{k}}=\langle\hat{a}_{\vec{j}}^{\dagger}\hat{a}_{\vec{k}}^{\nodagger}\rangle.
\end{equation}
However, it was shown by Read in Ref.~\cite{Read1989} that a specific type of ``dressed'' CBs can exhibit long-range order.
Extending this approach from the continuum to the lattice, we perform the nonlocal, non-unitary transformation
\begin{equation}\label{eq:read_cb}
    \hat{b}^{\dagger}_{\mathrm{R},\vec{j}}\equiv \hat{a}^{\dagger}_{\vec{j}} \left[ \hat{U}(z_{\vec{j}})\right]^m=\hat{a}^{\dagger}_{\vec{j}}\prod_{\vec{k}\neq \vec{j}}(z_{\vec{j}}-z_{\vec{k}})^{m\hat{n}_{\vec{k}}},
\end{equation}
to define the CB operators $\hat{b}^{(\dagger)}_{\mathrm{R}, \vec{j}}$.
Here, $\hat{U}(z_{\vec{j}})$ is the lattice analog of Laughlin's quasihole operator.
The transformation in \cref{eq:read_cb} attaches $m$ quasiholes (vortices) to each particle [\cref{fig:Sketch}(b)], rendering the resulting composite object strictly bosonic~\cite{SuppMat}.
We note that this transformation is not unitary and therefore any correlators obtained from the CB operators have to be properly normalized [see below].
States from the Laughlin sequence at filling factor $\nu \!=\! \nicefrac{1}{m}$ are believed to form a fluid of such Read-CBs that undergo Bose-Einstein condensation~\cite{Rezayi1988,Read1989}.
Consequently, the CB correlation function exhibits off-diagonal long-range order [\cref{fig:Sketch}(c)],
\begin{equation}\label{eq:read_corr}
    \rho^{\rm (R)}_{\vec{j},\vec{k}}=\langle\hat{b}_{\mathrm{R},\vec{j}}^{\dagger}\hat{b}_{\mathrm{R},\vec{k}}^{\nodagger}\rangle \xrightarrow{|\vec{j}-\vec{k}|\rightarrow \infty}\bar{n}.
\end{equation}

Analogously, a similar transformation can be performed for the particle-flux composita suggested by Girvin~and~MacDonald~\cite{Girvin1987}.
Here, the $^m$CBs, $\hat{b}_{\mathrm{GM}; \vec{i}}^{(\dagger)}$, are formed by attaching $m$ flux tubes to every particle:
\begin{equation}
    \hbd_{\mathrm{GM}; \vec{j}} \equiv \had_{\vec{j}} \de^{\di m \sum_{k} \arg\left(z_{\vec{j}} - z_{\vec{k}}\right) \hn_{\vec{k}}}.
\end{equation}
These composita in turn exhibit quasi-long range order with correlations decaying as [\cref{fig:Sketch}(c)]
\begin{equation}\label{eq:gm_corr}
    \rho^{\rm (GM)}_{\vec{j},\vec{k}}=\braket{\hbd_{\mathrm{GM}; \vec{j}} \hb_{\mathrm{GM}; \vec{k}}} \propto |\vec{j}-\vec{k}|^{-m/2}.
\end{equation}
\begin{figure}[t]
    \centering
    \includegraphics{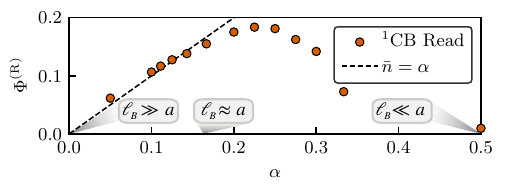}
    \caption{
       	Averaged two-point correlations $\Phi^{(\rm R)}$ [\cref{eq:order_parameter}] of Read-type CBs in the bulk of a non-interacting Chern insulator at $\nu\!=\!1$ for a system of size $L\times L \!=\! 10\times10$.
        The particle number is varied with $\alpha$ to keep the filling factor constant.
        The dashed line corresponds to the continuum prediction of the long-range behavior of the Read-CB correlator, $\lim_{r\rightarrow \infty}\rho^{(\rm R)}(r) \!=\! \alpha$.
        For small flux $\alpha \!\leq\! 0.2$ we find long-range order, $\Phi_{\rm R} \!\approx\! \bar{n} \!=\!\alpha$, whereas with increasing lattice effects the long-range correlations fade away.
    }
    \label{fig:Alphascan}
\end{figure}
\prlsection{Long-range order}
To establish the existence of topological CB long-range order in lattice systems, we first consider the non-interacting Chern insulator state at $\nu\!=\!1$ and investigate the role of lattice effects for a system of size $L \!=\! 10$ by tuning the flux per plaquette $\alpha$ from the Landau level limit ($\alpha \!\to\! 0$) to the highly dispersive lattice regime (large $\alpha \!\gtrsim\! \nicefrac{1}{5}$).
In~\cref{fig:Alphascan}, we monitor the averaged CB correlations in the bulk,
\begin{equation}
    \Phi^{(\rm R)} = \underset{|\vec{j}-\vec{k}|/\ell_B > 2}{\mathrm{avg}} |\rho_{\vec{j},\vec{k}}^{\rm (R)}|,
    \label{eq:order_parameter}
\end{equation}
where we discard the outermost sites at the edge of the system as well as short-distance correlations which are artificially suppressed by the quasihole formation [see below].
For the case of a non-interacting Chern insulator the composite objects consist of the constituent fermions with a single ($m\!=\!1$), integer-charged hole attached.
Crucially, as $\hat{U}(z_{\vec{j}})$ is non-unitary and creates a quasihole of radius $\sim\! m\ell_B$, when computing the Read correlation function the expectation values have to be normalized with respect to the hole density~\cite{SuppMat}.
While true long-range order with \mbox{$\Phi^{(\rm R)} \!\approx\! \bar{n} \!=\! \alpha$} emerges robustly over a wide range of small fluxes, strong lattice effects above $\alpha_c \!\approx\! \nicefrac{1}{5}$ affect the microscopic structure of the topological Chern insulator compared to its continuum analog, resulting in a sudden drop of $\Phi^{(\rm R)}$.
Explicitly taking into account the Bloch wave functions of the topological bands beyond simple Landau levels is expected to extend the simple quasi-continuum construction considered here to systems with strong lattice effects and opens further directions towards (fractional) Chern insulators in the absence of external magnetic fields~\cite{Cai2023,Zeng2023,Xu2023}.
To obtain robust topological states while also keeping some lattice effects, we will from now on fix the flux to $\alpha\!=\!\nicefrac{1}{6}$ such that $\ell_B \!\approx\! a$. 

\begin{figure*}[ht]
    \centering
    \includegraphics{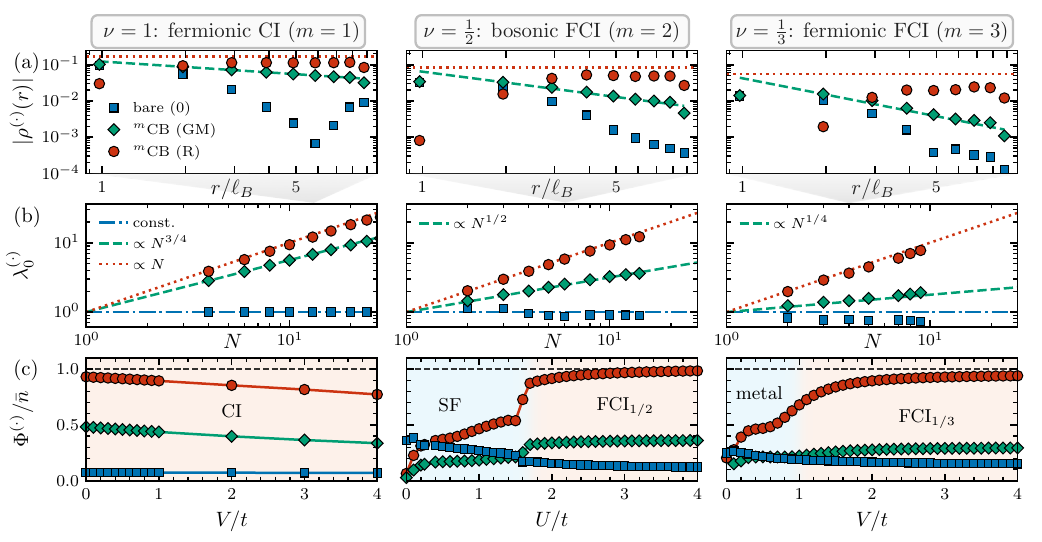}
    \caption{
       	(a) Two-point correlation functions $\rho^{(\cdot)}(r)$ of bare particles, Read (R), and Girvin-MacDonald (GM) CBs versus Euclidean distance $r/\ell_B$. 
        Columns display a noninteracting Chern insulator (left, $N\!=\!13$), a bosonic $\nicefrac{1}{2}$-Laughlin state (center, $N\!=\!6$), and a fermionic $\nicefrac{1}{3}$-Laughlin state (right, $N\!=\!4$). 
        The numerical data agrees well with the continuum predictions (dashed lines): true long-range correlations for the Read-CBs ($\rho^{\rm (R)}(r) \!\approx\! \bar{n} \!=\! \nicefrac{\alpha}{m}$ [\cref{eq:read_corr}]; red) and an algebraic decay for GM-CBs ($\rho^{(\rm GM)}(r) \!\propto\! r^{-m/2}$ [\cref{eq:gm_corr}]; green).
        (b) This behavior is further confirmed by the particle-number scaling of the leading eigenvalue $\lambda^{(\cdot)}_{0}$ of the two-point correlation matrix for the states in (a), evaluated at a fixed magnetic filling $\nu\!=\!\nicefrac{1}{m}$. 
        The lines indicate true condensation ($\propto \! N$, red, Read-CBs), quasi-condensation ($\propto \! N^{\nicefrac{(4-m)}{4}}$, green, GM-CBs), and absence of condensation (constant, blue, bare particles).
        (c) The order parameters $\Phi^{(\cdot)}/\bar{n}$ for the different CB constructions reveal phase transitions between trivial (blue shading) and topological phases (red shading) as the interaction strength is varied (nearest-neighbor $V$ for fermions, on-site $U$ for soft-core bosons with $n_{\rm max}\!=\!3$). 
        In all panels, the flux per plaquette is $\alpha \!=\! \nicefrac{1}{6}$, in (a)~and~(c) the system is of size $L\times L\!=\!10\times 10$.
    }
    \label{fig:CombinedResults}
\end{figure*}

Before turning towards more complex fractional Chern insulators, we briefly comment on the spatial structure of the correlations in the left panel of~\cref{fig:CombinedResults}(a).
In particular, we fix a reference site $\vec{j}$ at the edge of the system and systematically vary the second site $\vec{k}$ to evaluate $\rho^{(\cdot)}(r\!=\!|\vec{j}-\vec{k}|)$ for the different correlators.
While the bare correlator $\rho^{(0)}_{\vec{j},\vec{k}} \!=\! \braket{\had_{\vec{j}}\ha_{\vec{k}}}$ decays exponentially (before its revival close to the opposite edge of the system), the Read-CB correlator $\rho_{\vec{j},\vec{k}}^{\rm (R)} \!=\! \braket{\hbd_{\vec{j}}\hb_{\vec{k}}}$ exhibits a depleted short-distance correlation hole due to the attached quasiholes before saturating to the expected macroscopic plateau $\rho^{\rm (R)}(r\to L) \!\approx\! \bar{n}$.
Moving beyond the non-interacting limit, we next consider the strongly correlated bosonic ($\nu\!=\!\nicefrac{1}{2}$) and fermionic ($\nu\!=\!\nicefrac{1}{3}$) Laughlin states, conclusively demonstrating the emergence of long-range order in the FQH regime.
For the bosonic case, we probe the ground state of $\hcH+\hcH_{\rm B}$ [\cref{eq:model,eq:boson_model}] for $L\!=\!10$ and $N\!=\!6 \!\approx\! \nicefrac{\alpha (L-1)^2}{2}$ in the hard-core boson limit ($U/t \!\rightarrow\! \infty$) to rigorously isolate the Laughlin state~\cite{Soerensen2005}.
By attaching $m\!=\!2$ quasiholes to each boson, we find topological long-range order [middle panel of~\cref{fig:CombinedResults}(a)].
For the fermionic case at the same system size, we compute the ground state of $\hcH+\hcH_{\rm F}$ [\cref{eq:model,eq:fermion_model}] with $N\!=\!4,~L\!=\!10$ and strong interaction, $V/t\!=\!10$~\cite{Motruk2017}.
This ground state correspondingly exhibits $m\!=\!3$ topological long-range order [right panel of~\cref{fig:CombinedResults}(a)].
Moreover, across all three scenarios, the emergence of quasi-long-range order of GM-CBs is confirmed in agreement with earlier studies~\cite{Pauw2024}.

We emphasize that the correlator $\rho_{\vec{j},\vec{k}}^{(R)}$ can be obtained from site-resolved measurements of densities and a single coherence in the basis of the constituent particles, $\hat{a}_{\vec{j}}^{(\dagger)}$.
An experimental protocol to extract such combined information was previously proposed in the context of GM-CBs~\cite{Pauw2024}, and can be directly extended to the present case [see End Matter].
Therefore, our findings not only demonstrate the existence of true macroscopic long-range order in discrete lattice systems, but also establish this correlator as a direct diagnostic marker for topological order in quantum-simulation experiments.

\prlsection{Composite boson condensation}
Having established the emergence of true long-range order in (fractional) Chern insulators, we now turn to the related question of CB condensation in such states.
This condensation can be revealed via the largest eigenvalue $\lambda^{(\mathrm{R})}_0$ of the correlator $\rho^{\rm (R)}_{\vec{j},\vec{k}}$.
As argued by Penrose and Onsager~\cite{Penrose1956,Yang1962}, in a condensate this largest eigenvalue grows linearly with the number of particles in the system, $\lambda^{(\mathrm{R})}_0 \!\propto\! N$, in agreement with a finite condensate fraction in the thermodynamic limit, \mbox{$\lim_{N\to\infty} \nicefrac{\lambda^{(\mathrm{R})}_0}{N} \!=\! \text{const}$}.
In contrast, due to the suppression of two-point correlations for the consituent particles, their correlator does not feature a dominant eigenvalue, resulting in \mbox{$\lambda^{(0)}_0 \!\approx\! \lambda^{(0)}_k \!\propto\! \mathcal{O}(1)$} for all eigenvalues $\lambda^{(0)}_k$.

By systematically increasing the system size $L$ and the particle number $N$ while keeping the flux per plaquette fixed at $\alpha\!=\!\nicefrac{1}{6}$, we can explore this scaling for the lattice states discussed above.
In all cases -- non-interacting Chern insulator at $\nu\!=\!1$, Laughlin states at $\nu\!=\!\nicefrac{1}{2}$ (bosons) and $\nicefrac{1}{3}$ (fermions) -- we find the expected scaling $\lambda^{(\mathrm{R})}_0 \!\propto\! N$ over a large range of particle numbers~[\cref{fig:CombinedResults}(b)].
Analogously, the GM-CBs can form a quasi-condensate with, for \mbox{$m\!\leq\!4$}, a diverging leading eigenvalue of $\rho^{(\mathrm{GM})}_{\vec{j},\vec{k}}$ scaling as $\lambda^{(\mathrm{GM})}_0 \!\propto\! N^{\nicefrac{(4-m)}{4}}$.
Indeed, we find this scaling for all lattice states investigated here [\cref{fig:CombinedResults}(b)], thereby directly confirming another prediction of the CB theory for lattice systems beyond our earlier studies of the long-range correlations~\cite{Pauw2024,Pauw2026}.

\prlsection{Resolving phase transitions}
The condensation of the Read-CBs suggests to use the bulk behavior of the Read-type correlator 
\begin{equation}
	    \Phi^{(\rm R)} = \underset{|\vec{j}-\vec{k}|/\ell_B > 2}{\mathrm{avg}} |\rho_{\vec{j},\vec{k}}^{\rm (R)}| = \begin{cases}
	    	\bar{n} & \text{topological order,} \\
	    	0 & \text{trivial,}
	    \end{cases}
\end{equation}
to define a nonlocal order parameter reminiscent of the corresponding bare order parameter routinely measured to resolve phase transitions between superfluids and Mott insulators~\cite{Greiner2002,Bloch2008}.

To demonstrate the diagnostic power of $\Phi^{(\rm R)}$, we track the quantum phase transitions across three distinct regimes of the interacting Hofstadter model. 
First, we consider the fermionic Chern insulator at $\nu\!=\!1$ in the presence of nearest-neighbor repulsion of increasing strength $V/t$~[\cref{eq:fermion_model}]. 
Since this state does not rely on interactions, but is instead a topological band insulator, it exhibits robust nonlocal order irrespective of the interactions, indicated by a finite $\Phi^{\rm (R)} \!\approx\! \bar{n}$ for all $V/t$ [left panel of~\cref{fig:CombinedResults}(c)].
Next, we consider the bosonic Hofstadter model at $\nu\!=\!\nicefrac{1}{2}$ where we introduce a soft-core cut-off of up to $n_{\rm {max}}\!=\!3$ bosons per site.
Tuning the on-site repulsion $U/t$ drives a phase transition from a topologically trivial symmetry-broken superfluid (SF) at low $U/t$ into the topologically ordered Laughlin state~\cite{Barkeshli2014,Barkeshli2015,Song2024,Lu2026,Lotri2026,Wang2026}. 
As shown in the middle panel of~\cref{fig:CombinedResults}(c), $\Phi^{(\mathrm{R})}$ successfully captures this transition: it is suppressed in the SF phase -- where local density fluctuations scramble the nonlocal order -- and establishes a finite plateau near $\Phi^{(\rm R)} \!\approx\! \bar{n} \!=\! \nicefrac{\alpha}{2}$ in the strongly interacting FQH regime.
Finally, for interacting fermions at $\nu \!=\! \nicefrac{1}{3}$, by tuning the nearest-neighbor repulsion $V/t$, we traverse the transition from a trivial metallic state into the topological Laughlin phase~\cite{Pauw2026}.
Once again, the onset of topological order is unambiguously resolved by the emergence of a CB condensate with the order parameter plateauing around $\bar{n} \!=\! \nicefrac{\alpha}{3}$ [right panel of \cref{fig:CombinedResults}(c)].
Although finite-size effects inherently broaden the expected sharp transition to a non-zero $\Phi^{(\rm R)}$, this effect systematically diminishes as the system size increases~\cite{SuppMat}.
While the system sizes studied here do not allow for a precise resolution of the the critical interaction strength, our data is consistent with earlier studies~\cite{Wang2026,Pauw2026}.

\prlsection{Discussion}
In this Letter, we have demonstrated the emergence of long-range correlations in topologically ordered lattice states, described by a composite boson condensate.
By extending Read's non-unitary quasihole transformation to lattices and explicitly evaluating the resulting bosonic correlators using MPS, we have established a nonlocal order parameter that captures the topological nature of the insulating bulk and serves as a practical diagnostic for quantum phase transitions between trivial and topological phases.

We emphasize that all observables presented in this Letter can be obtained directly from local snapshot measurements [see End Matter], which are routinely available in quantum-simulation experiments~\cite{Weitenberg2026}.
Accordingly, our results are immediately applicable to state-of-the-art quantum simulators, for example with ultracold atoms and superconducting qubits, where the nonlocal order parameter offers a way to certify desired target states.
In this context, future studies should further clarify the robustness of the composite boson condensate and its exotic order at finite temperature and in the presence of disorder and imperfections during (quasi-)adiabatic state preparation~\cite{Leonard2023}.
While our current study considered single-component, Abelian quantum Hall states, this formalism naturally extends to more complex phases, such as multi-component Halperin states~\cite{Halperin1983,Halperin1984}, or non-Abelian topological states, such as the $p$-wave paired Pfaffian state introduced by Moore~and~Read~\cite{Moore1991}.
In this context, it would also be interesting to extend the analysis performed here to fractional Chern insulators in weak and vanishing magnetic fields~\cite{Cai2023,Zeng2023,Xu2023}, where the flux attachment construction of Girvin~and~MacDonald becomes physically ambiguous, whereas Read's quasihole attachment remains robust and rigorously defined.
Along this line, it is also interesting to rephrase the arguments used here in terms of orbital attachment~\cite{Haldane2011,Haldane2018a}.
Moreover, systematically truncating the spatial support of the nonlocal operators would enable implementations of our approach in the thermodynamic limit where evaluating intrinsically global objects is not possible.
In this way, the results presented here are, among others, relevant to solid-state systems and numerical simulations using uniform tensor networks, thereby offering a powerful diagnostic for strongly correlated topological matter.

\prlsection{Acknowledgments}
The authors thank A. Eckardt and T. Roscilde for discussions and F. Grusdt for collaboration on earlier related work.
All DMRG simulations were performed with the \textsc{SyTen}-toolkit~\cite{Syten,Hubig17}.
F.J.P. acknowledges support by the Deutsche Forschungsgemeinschaft (DFG, German Research Foundation) under Germany’s Excellence Strategy-426 EXC-2111-390814868.
Work in Brussels was supported by the ERC Grant LATIS, the FRS-FNRS (Belgium), the EOS project CHEQS, and the Fondation ULB.

\bibliography{literature.bib}

@Article{Tsui1982,
  author     = {D. C. Tsui and H. L. Stormer and A. C. Gossard},
  journal    = {Physical Review Letters},
  title      = {{Two-Dimensional Magnetotransport in the Extreme Quantum Limit}},
  year       = {1982},
  month      = {may},
  number     = {22},
  pages      = {1559--1562},
  volume     = {48},
  doi        = {10.1103/physrevlett.48.1559},
}

@article{White1992,
	title = {Density matrix formulation for quantum renormalization groups},
	author = {White, Steven R.},
	journal = {Phys. Rev. Lett.},
	volume = {69},
	issue = {19},
	pages = {2863--2866},
	numpages = {0},
	year = {1992},
	month = {Nov},
	publisher = {American Physical Society},
	doi = {10.1103/PhysRevLett.69.2863},
	url = {https://link.aps.org/doi/10.1103/PhysRevLett.69.2863}
}

@Article{Read1989,
	author     = {N. Read},
	journal    = {Physical Review Letters},
	title      = {{Order Parameter and Ginzburg-Landau Theory for the Fractional Quantum Hall Effect}},
	year       = {1989},
	month      = {jan},
	number     = {1},
	pages      = {86--89},
	volume     = {62},
	doi        = {10.1103/physrevlett.62.86},
}

@InBook{Girvin1988,
	author     = {Girvin, S. M.},
	editor     = {Leavens, C. Richard and Taylor, Roger},
	pages      = {333--346},
	publisher  = {Springer US},
	title      = {{Off-Diagonal Long-Range Order in the Quantum Hall Effect}},
	year       = {1988},
	address    = {Boston, MA},
	isbn       = {978-1-4613-1045-7},
	volume     = {179},
	booktitle  = {Interfaces, Quantum Wells, and Superlattices},
	doi        = {10.1007/978-1-4613-1045-7_18},
}

@Article{Laughlin1983,
	author     = {R. B. Laughlin},
	journal    = {Physical Review Letters},
	title      = {{Anomalous Quantum Hall Effect: An Incompressible Quantum Fluid with Fractionally Charged Excitations}},
	year       = {1983},
	month      = {may},
	number     = {18},
	pages      = {1395--1398},
	volume     = {50},
	doi        = {10.1103/physrevlett.50.1395},
}

@Article{Palm2022,
	author        = {Felix A. Palm and Sam Mardazad and Annabelle Bohrdt and Ulrich Schollwöck and Fabian Grusdt},
	title         = {{Snapshot-based detection of $\ensuremath{\nu}=\frac{1}{2}$ Laughlin states: Coupled chains and central charge}},
	year          = {2022},
	month         = aug,
	journal = {Physical Review B},
	volume = {106},
	number = {8},
	pages = {L081108},
	doi = {10.1103/PhysRevB.106.L081108},
}

@Article{Soerensen2005,
	author     = {Anders S. S{\o}rensen and Eugene Demler and Mikhail D. Lukin},
	journal    = {Physical Review Letters},
	title      = {{Fractional Quantum Hall States of Atoms in Optical Lattices}},
	year       = {2005},
	month      = {mar},
	number     = {8},
	pages      = {086803},
	volume     = {94},
	doi        = {10.1103/physrevlett.94.086803},
}

@Article{Hafezi2007,
	author     = {M. Hafezi and A. S. S{\o}rensen and E. Demler and M. D. Lukin},
	journal    = {Physical Review A},
	title      = {{Fractional quantum Hall effect in optical lattices}},
	year       = {2007},
	month      = {aug},
	number     = {2},
	pages      = {023613},
	volume     = {76},
	doi        = {10.1103/physreva.76.023613},
}

@Article{Motruk2017,
	author    = {Johannes Motruk and Frank Pollmann},
	journal   = {Physical Review B},
	title     = {{Phase transitions and adiabatic preparation of a fractional Chern insulator in a boson cold-atom model}},
	year      = {2017},
	month     = {oct},
	number    = {16},
	pages     = {165107},
	volume    = {96},
	doi       = {10.1103/physrevb.96.165107},
}

@Article{Boesl2022,
	author        = {Julian Boesl and Rohit Dilip and Frank Pollmann and Michael Knap},
	journal       = {Physical Review B},
	title         = {{Characterizing fractional topological phases of lattice bosons near the first Mott lobe}},
	year          = {2022},
	month         = {feb},
	number        = {7},
	pages         = {075135},
	volume        = {105},
	doi           = {10.1103/physrevb.105.075135},
}

@Article{Wang2022,
	author        = {Botao Wang and Xiaoyu Dong and Andr{\'{e}} Eckardt},
	journal       = {{SciPost} Physics},
	title         = {{Measurable signatures of bosonic fractional Chern insulator states and their fractional excitations in a quantum-gas microscope}},
	year          = {2022},
	month         = nov,
	number        = {3},
	volume        = {12},
	url           = {10.21468/scipostphys.12.3.095},
}

@Article{Moore1991,
	author    = {Gregory Moore and Nicholas Read},
	journal   = {Nuclear Physics B},
	title     = {{Nonabelions in the fractional quantum hall effect}},
	year      = {1991},
	month     = {aug},
	number    = {2-3},
	pages     = {362--396},
	volume    = {360},
	doi       = {10.1016/0550-3213(91)90407-o},
}

@article{Schollwock2011,
	doi = {10.1016/j.aop.2010.09.012},
	
	url = {https://doi.org/10.1016%2Fj.aop.2010.09.012},
	
	year = 2011,
	month = {jan},
	
	publisher = {Elsevier {BV}
	},
	
	volume = {326},
	
	number = {1},
	
	pages = {96--192},
	
	author = {Ulrich Schollwöck},
	
	title = {The density-matrix renormalization group in the age of matrix product states},
	
	journal = {Annals of Physics}
}

@Article{Hubig2015,
	author    = {C. Hubig and I. P. McCulloch and U. Schollw\"ock and F. A. Wolf},
	journal   = {Physical Review B},
	title     = {Strictly single-site {DMRG} algorithm with subspace expansion},
	year      = {2015},
	month     = {apr},
	number    = {15},
	pages     = {155115},
	volume    = {91},
	doi       = {10.1103/physrevb.91.155115},
	publisher = {American Physical Society ({APS})},
}

@Misc{SyTen,
	author     = {Claudius Hubig and Felix Lachenmaier and Nils-Oliver Linden and Teresa Reinhard and Leo Stenzel and Andreas Swoboda and Martin Grundner and Sam Mardazad and Fabian Pauw and Sebastian Paeckel},
	title      = {The \textsc{SyTen} Toolkit},
	year       = {2023},
	url        = {https://syten.eu},
}

@article{Girvin1987,
	title = {{Off-diagonal long-range order, oblique confinement, and the fractional quantum Hall effect}},
	author = {Girvin, S. M. and MacDonald, A. H.},
	journal = {Phys. Rev. Lett.},
	volume = {58},
	issue = {12},
	pages = {1252--1255},
	numpages = {0},
	year = {1987},
	month = {Mar},
	publisher = {American Physical Society},
	doi = {10.1103/PhysRevLett.58.1252},
	url = {https://link.aps.org/doi/10.1103/PhysRevLett.58.1252}
}

@article{Rezayi1988,
	title = {{Off-Diagonal Long-Range Order in Fractional Quantum-Hall-Effect States}},
	author = {Rezayi, E. H. and Haldane, F. D. M.},
	journal = {Phys. Rev. Lett.},
	volume = {61},
	issue = {17},
	pages = {1985--1988},
	numpages = {0},
	year = {1988},
	month = {Oct},
	publisher = {American Physical Society},
	doi = {10.1103/PhysRevLett.61.1985},
	url = {https://link.aps.org/doi/10.1103/PhysRevLett.61.1985}
}

@Article{Leonard2023,
	author = {L{\'e}onard, Julian and Kim, Sooshin and Kwan, Joyce and Segura, Perrin and Grusdt, Fabian and Repellin, C{\'e}cile and Goldman, Nathan and Greiner, Markus},
	date = {2023/07/01},
	doi = {10.1038/s41586-023-06122-4},
	id = {L{\'e}onard2023},
	isbn = {1476-4687},
	journal = {Nature},
	number = {7970},
	pages = {495--499},
	title = {{Realization of a fractional quantum Hall state with ultracold atoms}},
	url = {https://doi.org/10.1038/s41586-023-06122-4},
	volume = {619},
	year = {2023},
}

@article{Greiner2002,
	author = {Greiner, Markus and Mandel, Olaf and Esslinger, Tilman and H{\"a}nsch, Theodor W. and Bloch, Immanuel},
	date = {2002/01/01},
	doi = {10.1038/415039a},
	id = {Greiner2002},
	isbn = {1476-4687},
	journal = {Nature},
	number = {6867},
	pages = {39--44},
	title = {{Quantum phase transition from a superfluid to a Mott insulator in a gas of ultracold atoms}},
	url = {https://doi.org/10.1038/415039a},
	volume = {415},
	year = {2002}}

@Article{Xu2023,
	author    = {Xu, Fan and Sun, Zheng and Jia, Tongtong and Liu, Chang and Xu, Cheng and Li, Chushan and Gu, Yu and Watanabe, Kenji and Taniguchi, Takashi and Tong, Bingbing and Jia, Jinfeng and Shi, Zhiwen and Jiang, Shengwei and Zhang, Yang and Liu, Xiaoxue and Li, Tingxin},
	journal   = {Physical Review X},
	title     = {{Observation of Integer and Fractional Quantum Anomalous Hall Effects in Twisted Bilayer MoTe$_2$}},
	year      = {2023},
	issn      = {2160-3308},
	month     = sep,
	number    = {3},
	pages     = {031037},
	volume    = {13},
	doi       = {10.1103/physrevx.13.031037},
}

@article{Pauw2024,
	title = {{Detecting hidden order in fractional Chern insulators}},
	author = {Pauw, F. J. and Palm, F. A. and Schollw\"ock, U. and Bohrdt, A. and Paeckel, S. and Grusdt, F.},
	journal = {Phys. Rev. Res.},
	volume = {6},
	issue = {2},
	pages = {023180},
	numpages = {18},
	year = {2024},
	month = {May},
	publisher = {American Physical Society},
	doi = {10.1103/PhysRevResearch.6.023180},
	url = {https://link.aps.org/doi/10.1103/PhysRevResearch.6.023180}
}

@phdthesis{Hubig17,
    author = {Claudius Hubig},
    title = {Symmetry-Protected Tensor Networks},
    school = {LMU München},
    year = {2017},
    url = {https://edoc.ub.uni-muenchen.de/21348/},
}

@article{Schollwock2005,
  title = {The density-matrix renormalization group},
  author = {Schollw\"ock, U.},
  journal = {Rev. Mod. Phys.},
  volume = {77},
  issue = {1},
  pages = {259--315},
  numpages = {0},
  year = {2005},
  month = {Apr},
  publisher = {American Physical Society},
  doi = {10.1103/RevModPhys.77.259},
  url = {https://link.aps.org/doi/10.1103/RevModPhys.77.259}
}

@article{Gerster2014,
  title = {Unconstrained tree tensor network: An adaptive gauge picture for enhanced performance},
  author = {Gerster, M. and Silvi, P. and Rizzi, M. and Fazio, R. and Calarco, T. and Montangero, S.},
  journal = {Phys. Rev. B},
  volume = {90},
  issue = {12},
  pages = {125154},
  numpages = {10},
  year = {2014},
  month = {Sep},
  publisher = {American Physical Society},
  doi = {10.1103/PhysRevB.90.125154},
  url = {https://link.aps.org/doi/10.1103/PhysRevB.90.125154}
}

@article{Penrose1956,
  title = {{Bose-Einstein Condensation and Liquid Helium}},
  author = {Penrose, Oliver and Onsager, Lars},
  journal = {Phys. Rev.},
  volume = {104},
  issue = {3},
  pages = {576--584},
  numpages = {0},
  year = {1956},
  month = {Nov},
  publisher = {American Physical Society},
  doi = {10.1103/PhysRev.104.576},
  url = {https://link.aps.org/doi/10.1103/PhysRev.104.576}
}

@article{Pauw2026,
  title = {{From hidden order to skyrmions: Quantum Hall states in an extended Hofstadter-Fermi-Hubbard model}},
  author = {Pauw, F. J. and Schollw\"ock, U. and Goldman, N. and Paeckel, S. and Palm, F. A.},
  journal = {Phys. Rev. B},
  volume = {113},
  issue = {3},
  pages = {035142},
  numpages = {16},
  year = {2026},
  month = {Jan},
  publisher = {American Physical Society},
  doi = {10.1103/cqt2-b327},
  url = {https://link.aps.org/doi/10.1103/cqt2-b327}
}

@article{Zhang2023,
  title = {{From frustration-free parent Hamiltonians to off-diagonal long-range order: Moore-Read and related states in second quantization}},
  author = {Zhang, Fanmao and Schossler, Matheus and Seidel, Alexander and Chen, Li},
  journal = {Phys. Rev. B},
  volume = {108},
  issue = {7},
  pages = {075142},
  numpages = {18},
  year = {2023},
  month = {Aug},
  publisher = {American Physical Society},
  doi = {10.1103/PhysRevB.108.075142},
  url = {https://link.aps.org/doi/10.1103/PhysRevB.108.075142}
}

@article{Chen2019,
  title = {Composite fermions in Fock space: Operator algebra, recursion relations, and order parameters},
  author = {Chen, Li and Bandyopadhyay, Sumanta and Yang, Kun and Seidel, Alexander},
  journal = {Phys. Rev. B},
  volume = {100},
  issue = {4},
  pages = {045136},
  numpages = {18},
  year = {2019},
  month = {Jul},
  publisher = {American Physical Society},
  doi = {10.1103/PhysRevB.100.045136},
  url = {https://link.aps.org/doi/10.1103/PhysRevB.100.045136}
}

@article{Mazaheri2015,
  title = {{Zero modes, bosonization, and topological quantum order: The Laughlin state in second quantization}},
  author = {Mazaheri, Tahereh and Ortiz, Gerardo and Nussinov, Zohar and Seidel, Alexander},
  journal = {Phys. Rev. B},
  volume = {91},
  issue = {8},
  pages = {085115},
  numpages = {12},
  year = {2015},
  month = {Feb},
  publisher = {American Physical Society},
  doi = {10.1103/PhysRevB.91.085115},
  url = {https://link.aps.org/doi/10.1103/PhysRevB.91.085115}
}

@Article{Wang2024,
  author        = {Wang, Can and Liu, Feng-Ming and Chen, Ming-Cheng and Chen, He and Zhao, Xian-He and Ying, Chong and Shang, Zhong-Xia and Wang, Jian-Wen and Huo, Yong-Heng and Peng, Cheng-Zhi and Zhu, Xiaobo and Lu, Chao-Yang and Pan, Jian-Wei},
  journal       = {Science},
  title         = {{Realization of fractional quantum Hall state with interacting photons}},
  year          = {2024},
  issn          = {1095-9203},
  month         = jan,
  number        = {6695},
  pages         = {579--584},
  volume        = {384},
  doi           = {10.1126/science.ado3912},
}

@article{Yang1962,
  title = {{Concept of Off-Diagonal Long-Range Order and the Quantum Phases of Liquid He and of Superconductors}},
  author = {Yang, C. N.},
  journal = {Rev. Mod. Phys.},
  volume = {34},
  issue = {4},
  pages = {694--704},
  numpages = {0},
  year = {1962},
  month = {Oct},
  publisher = {American Physical Society},
  doi = {10.1103/RevModPhys.34.694},
  url = {https://link.aps.org/doi/10.1103/RevModPhys.34.694}
}

@Article{Landau1937,
  author       = {L. D. Landau},
  journal      = {Zh. Eksp. Teor. Fiz.},
  title        = {On the theory of phase transitions},
  year         = {1937},
  pages        = {19--32},
  volume       = {7},
  doi          = {10.1016/b978-0-08-010586-4.50034-1},
}

@Article{Ginsburg1950,
  author       = {V. L. Ginsburg and L. D. Landau},
  journal      = {Zh. Eksp. Teor. Fiz.},
  title        = {On the theory of superconductivity},
  year         = {1950},
  pages        = {1064--1082},
  volume       = {20},
  doi          = {10.1016/b978-0-08-010586-4.50078-x},
}

@article{Wegner1971,
    author = {Wegner, Franz J.},
    title = {{Duality in Generalized Ising Models and Phase Transitions without Local Order Parameters}},
    journal = {Journal of Mathematical Physics},
    volume = {12},
    number = {10},
    pages = {2259-2272},
    year = {1971},
    month = {10},
    doi = {10.1063/1.1665530},
    url = {https://doi.org/10.1063/1.1665530},
}

@article{Wilson1974,
  title = {Confinement of quarks},
  author = {Wilson, Kenneth G.},
  journal = {Phys. Rev. D},
  volume = {10},
  issue = {8},
  pages = {2445--2459},
  numpages = {0},
  year = {1974},
  month = {Oct},
  publisher = {American Physical Society},
  doi = {10.1103/PhysRevD.10.2445},
  url = {https://link.aps.org/doi/10.1103/PhysRevD.10.2445}
}

@article{denNijs1989,
  title = {Preroughening transitions in crystal surfaces and valence-bond phases in quantum spin chains},
  author = {den Nijs, Marcel and Rommelse, Koos},
  journal = {Phys. Rev. B},
  volume = {40},
  issue = {7},
  pages = {4709--4734},
  numpages = {0},
  year = {1989},
  month = {Sep},
  publisher = {American Physical Society},
  doi = {10.1103/PhysRevB.40.4709},
  url = {https://link.aps.org/doi/10.1103/PhysRevB.40.4709}
}

@article{DallaTorre2006,
  title = {{Hidden Order in 1D Bose Insulators}},
  author = {Dalla Torre, Emanuele G. and Berg, Erez and Altman, Ehud},
  journal = {Phys. Rev. Lett.},
  volume = {97},
  issue = {26},
  pages = {260401},
  numpages = {4},
  year = {2006},
  month = {Dec},
  publisher = {American Physical Society},
  doi = {10.1103/PhysRevLett.97.260401},
  url = {https://link.aps.org/doi/10.1103/PhysRevLett.97.260401}
}

@article{Kennedy1992,
  title = {{Hidden ${\mathrm{Z}}_{2}$\ifmmode\times\else\texttimes\fi{}${\mathrm{Z}}_{2}$ symmetry breaking in Haldane-gap antiferromagnets}},
  author = {Kennedy, Tom and Tasaki, Hal},
  journal = {Phys. Rev. B},
  volume = {45},
  issue = {1},
  pages = {304--307},
  numpages = {0},
  year = {1992},
  month = {Jan},
  publisher = {American Physical Society},
  doi = {10.1103/PhysRevB.45.304},
  url = {https://link.aps.org/doi/10.1103/PhysRevB.45.304}
}

@article{Fredenhagen1983,
	author = {Fredenhagen, Klaus and Marcu, Mihail},
	date = {1983/03/01},
	doi = {10.1007/BF01206315},
	id = {Fredenhagen1983},
	isbn = {1432-0916},
	journal = {Communications in Mathematical Physics},
	number = {1},
	pages = {81--119},
	title = {{Charged states in $\mathbb{Z}_2$ gauge theories}},
	url = {https://doi.org/10.1007/BF01206315},
	volume = {92},
	year = {1983}}

@misc{Wang2026,
      title={{Emergent QED$_3$ at the bosonic Laughlin state to superfluid transition}}, 
      author={Taige Wang and Xue-Yang Song and Michael P. Zaletel and T. Senthil},
      year={2025},
      eprint={2507.07611},
      archivePrefix={arXiv},
      primaryClass={cond-mat.str-el},
      url={https://arxiv.org/abs/2507.07611}, 
}

@article{Lotri2026,
  title = {{Paired Parton Trial States for the Superfluid-Fractional Chern Insulator Transition}},
  author = {Lotri\v{c}, T. and Simon, S. H.},
  journal = {Phys. Rev. Lett.},
  volume = {136},
  issue = {9},
  pages = {096601},
  numpages = {6},
  year = {2026},
  month = {Mar},
  publisher = {American Physical Society},
  doi = {10.1103/l9rp-y68y},
  url = {https://link.aps.org/doi/10.1103/l9rp-y68y}
}

@article{Song2024,
   author={Song, Xue-Yang and Zhang, Ya-Hui and Senthil, T.},
   title={{Phase transitions out of quantum Hall states in moiré materials}},
   journal={Physical Review B},
   volume={109},
   number={8},
   pages={085143},
   year={2024},
   month={2},
   publisher={American Physical Society},
   doi={10.1103/PhysRevB.109.085143}
}

@article{Lu2026,
   author={Lu, Hongyu and Wu, Han-Qing and Chen, Bin-Bin and Meng, Zi Yang},
   title={{Vestigial gapless boson density wave emerging between fractional Chern insulator and finite-momentum supersolid}},
   journal={Physical Review B},
   volume={113},
   number={3},
   pages={035141},
   year={2026},
   month={1},
   publisher={American Physical Society},
   doi={10.1103/PhysRevB.113.035141}
}

@article{Barkeshli2015,
   author={Barkeshli, M. and Yao, N. Y. and Laumann, C. R.},
   title={{Continuous Preparation of a Fractional Chern Insulator}},
   journal={Physical Review Letters},
   volume={115},
   number={2},
   pages={026802},
   year={2015},
   month={7},
   publisher={American Physical Society},
   doi={10.1103/PhysRevLett.115.026802}
}

@Article{Halperin1983,
  author    = {Halperin, B. I.},
  journal   = {Helvetica Physica Acta},
  title     = {{Theory of the quantized Hall conductance}},
  year      = {1983},
  volume    = {56},
  doi       = {10.5169/SEALS-115362},
}

@article{Halperin1984,
  title = {{Statistics of Quasiparticles and the Hierarchy of Fractional Quantized Hall States}},
  author = {Halperin, B. I.},
  journal = {Phys. Rev. Lett.},
  volume = {52},
  issue = {18},
  pages = {1583--1586},
  numpages = {0},
  year = {1984},
  month = {Apr},
  publisher = {American Physical Society},
  doi = {10.1103/PhysRevLett.52.1583},
  url = {https://link.aps.org/doi/10.1103/PhysRevLett.52.1583}
}

@misc{SuppMat,
  note = {See Supplemental Material, containing Ref.~\cite{Hubig2015}, for numerical details and additional supporting data.}
}

@Article{Cai2023,
	author       = {Cai, Jiaqi and Anderson, Eric and Wang, Chong and Zhang, Xiaowei and Liu, Xiaoyu and Holtzmann, William and Zhang, Yinong and Fan, Fengren and Taniguchi, Takashi and Watanabe, Kenji and Ran, Ying and Cao, Ting and Fu, Liang and Xiao, Di and Yao, Wang and Xu, Xiaodong},
	journal      = {Nature},
	title        = {{Signatures of fractional quantum anomalous Hall states in twisted MoTe2}},
	year         = {2023},
	issn         = {1476-4687},
	month        = jun,
	number       = {7981},
	pages        = {63--68},
	volume       = {622},
	doi          = {10.1038/s41586-023-06289-w},
}

@Article{Zeng2023,
	author       = {Zeng, Yihang and Xia, Zhengchao and Kang, Kaifei and Zhu, Jiacheng and Knüppel, Patrick and Vaswani, Chirag and Watanabe, Kenji and Taniguchi, Takashi and Mak, Kin Fai and Shan, Jie},
	journal      = {Nature},
	title        = {{Thermodynamic evidence of fractional Chern insulator in moiré MoTe2}},
	year         = {2023},
	issn         = {1476-4687},
	month        = jul,
	number       = {7981},
	pages        = {69--73},
	volume       = {622},
	doi          = {10.1038/s41586-023-06452-3},
}

@article{Bloch2008,
  title = {Many-body physics with ultracold gases},
  author = {Bloch, Immanuel and Dalibard, Jean and Zwerger, Wilhelm},
  journal = {Rev. Mod. Phys.},
  volume = {80},
  issue = {3},
  pages = {885--964},
  year = {2008},
  publisher = {American Physical Society},
  doi = {10.1103/RevModPhys.80.885}
}

@Article{Haldane2011,
	author       = {Haldane, F. D. M.},
	journal      = {Physical Review Letters},
	title        = {Geometrical {Description} of the {Fractional} {Quantum} {Hall} {Effect}},
	year         = {2011},
	number       = {11},
	pages        = {116801},
	volume       = {107},
	doi          = {10.1103/PhysRevLett.107.116801},
	publisher    = {American Physical Society},
}

@InBook{Haldane2018a,
	author       = {Haldane, F. Duncan M.},
	publisher    = {WORLD SCIENTIFIC},
	title        = {Geometry of flux attachment in the fractional quantum hall effect states},
	year         = {2018},
	isbn         = {9789813271340},
	month        = Aug,
    pages        = {2-2},
    chapter      = {},
	booktitle    = {Topological Phase Transitions and New Developments},
	doi          = {10.1142/9789813271340_0002},
}

@misc{Weitenberg2026,
      title={Protocols for a many-body phase microscope: From coherences and d-wave superconductivity to Green's functions}, 
      author={Christof Weitenberg and Luca Asteria and Ola Carlsson and Annabelle Bohrdt and Fabian Grusdt},
      year={2026},
      eprint={2602.12142},
      archivePrefix={arXiv},
      primaryClass={cond-mat.quant-gas},
      url={https://arxiv.org/abs/2602.12142}, 
}

@article{McGreevy2023,
   title={Generalized Symmetries in Condensed Matter},
   volume={14},
   ISSN={1947-5462},
   url={http://dx.doi.org/10.1146/annurev-conmatphys-040721-021029},
   DOI={10.1146/annurev-conmatphys-040721-021029},
   number={1},
   journal={Annual Review of Condensed Matter Physics},
   publisher={Annual Reviews},
   author={McGreevy, John},
   year={2023},
   month=Mar, pages={57–82} }

@article{Gaiotto2015,
   title={Generalized global symmetries},
   volume={2015},
   url={http://dx.doi.org/10.1007/JHEP02(2015)172},
   DOI={10.1007/jhep02(2015)172},
   number={2},
   pages={172},
   journal={Journal of High Energy Physics},
   publisher={Springer Science and Business Media LLC},
   author={Gaiotto, Davide and Kapustin, Anton and Seiberg, Nathan and Willett, Brian},
   year={2015},
   month=Feb }

@article{Pace2023,
  title = {Exact emergent higher-form symmetries in bosonic lattice models},
  author = {Pace, Salvatore D. and Wen, Xiao-Gang},
  journal = {Phys. Rev. B},
  volume = {108},
  issue = {19},
  pages = {195147},
  numpages = {32},
  year = {2023},
  month = {Nov},
  publisher = {American Physical Society},
  doi = {10.1103/PhysRevB.108.195147},
  url = {https://link.aps.org/doi/10.1103/PhysRevB.108.195147}
}

@article{Barkeshli2014,
  title = {Continuous transition between fractional quantum Hall and superfluid states},
  author = {Barkeshli, Maissam and McGreevy, John},
  journal = {Phys. Rev. B},
  volume = {89},
  issue = {23},
  pages = {235116},
  numpages = {6},
  year = {2014},
  month = {Jun},
  publisher = {American Physical Society},
  doi = {10.1103/PhysRevB.89.235116},
  url = {https://link.aps.org/doi/10.1103/PhysRevB.89.235116}
}

@Article{Impertro2024,
  author       = {Impertro, Alexander and Karch, Simon and Wienand, Julian F. and Huh, SeungJung and Schweizer, Christian and Bloch, Immanuel and Aidelsburger, Monika},
  journal      = {Physical Review Letters},
  title        = {{Local Readout and Control of Current and Kinetic Energy Operators in Optical Lattices}},
  year         = {2024},
  issn         = {1079-7114},
  month        = aug,
  number       = {6},
  pages        = {063401},
  volume       = {133},
  doi          = {10.1103/physrevlett.133.063401},
}

\onecolumngrid
\begin{center}
	{\bfseries End Matter}
\end{center}
\setcounter{section}{0} 
\renewcommand{\thesection}{\Alph{section}} 
\refstepcounter{section}

\twocolumngrid
\section{Experimentally accessing the Read-CB correlator}
\label{App:ReadCBMeasurement}
Here we briefly discuss how the experimental protocol introduced in Ref.~\cite{Pauw2024} for the GM-CBs can be extended directly to probe the Read-CB correlator. 
The primary distinction lies in the non-unitary nature of the Read quasihole operators $\hat{U}$, which necessitates a proper normalization of the correlator. 
Specifically, the normalized Read-CB correlator between two sites $\vec{j}$ and $\vec{k}$ is given by
\begin{equation} \label{Eq:NormalizedReadCB}
    \mathcal{C}^{(R)}_{\vec{j},\vec{k}} = \frac{\braket{\hat{\rho}^{(R)}_{\vec{j},{\vec{k}}}}}{\sqrt{\braket{\hat{U}^{\dagger}_{\vec{j}}\hat{U}^{\nodagger}_{\vec{j}}}\braket{\hat{U}^{\nodagger}_{\vec{k}}\hat{U}^{\dagger}_{\vec{k}}}}},
\end{equation}
where the numerator represents the unnormalized Read composite boson correlator, $\hat{\rho}^{(R)}_{\vec{j},{\vec{k}}} \!=\! \hat{a}_{\vec{j}}^{\dagger} \hat{U}_{\vec{j}}^{\dagger} \hat{U}_{\vec{k}}^{\nodagger} \hat{a}_{\vec{k}}^{\nodagger}$. 
Extracting this quantity requires the evaluation of three distinct expectation values, which can be obtained via independent sets of experimental snapshots from state-of-the-art quantum gas microscopes~\cite{Pauw2024,Weitenberg2026}.

\paragraph{Evaluating the numerator:}
The unnormalized correlator in the numerator is an off-diagonal observable that maps perfectly onto the synthetic bilayer framework introduced in Ref.~\cite{Pauw2024}.
Identifying the sites $\vec{j}$ and $\vec{k}$ with the internal states $(x,y,+)$ and $(x,y,-)$ respectively, the operator takes the form $\hat{a}_{x,y,+}^{\dagger} \hat{a}_{x,y,-}^{\nodagger} f^{(R)}(\{\hat{n}\})$.
Here, the function \mbox{$f^{(R)}(\{\hat{n}\}) \!=\! \hat{U}_{\vec{j}}^{\dagger} \hat{U}_{\vec{k}}^{\nodagger}$} encapsulates the specific phase and amplitude modifications of the Read quasihole attachment, which depends strictly on the site-local Fock spaces (density distribution) of the remaining lattice sites. 
The observable can be extracted using the following interferometric $\nicefrac{\pi}{2}$-pulse sequence, a variant of which has recently been implemented in Ref.~\cite{Impertro2024}.
First, by switching on an additional inter-layer coupling between the internal states $(x,y,\pm)$ for a time corresponding to a $\pi/2$-pulse followed by a local detuning, the bosonic operators are rotated to the new measurement  basis:
\begin{equation}
    \hat{a}^{\pm}_{x,y}(\varphi)=(\hat{a}_{x,y,+}\pm \de^{-\di\varphi} \hat{a}_{x,y,-})/\sqrt{2},
\end{equation}
where the phase-shift $\varphi$ is controlled by the detuning offset $\Delta$ and the detuning time $\tau$, i.e. $\varphi-\pi/2 \propto\Delta\tau$.
The occupation number operator in the rotated basis is then given by
\begin{equation}
    \hat{n}^{\pm}_{x,y}(\varphi)=(\hat{n}_{x,y,+}+\hat{n}_{x,y,-}\pm \hat{j}_{x,y}(\varphi))/\sqrt{2},
\end{equation}
where $\hat{j}_{x,y}(\varphi)=-\di \de^{\di \varphi}\hat{a}^{\dagger}_{x,y,-}\hat{a}^{\nodagger}_{x,y,+}+\rm{H.c.}$ is the generalized current operator.
Taking a snapshot in this rotated basis allows to read-off the occupation number in the original basis while the generalized current $\hat{j}_{x,y}$ can be resolved simultaneously for any value of the offset angle $\varphi$.
Combining this protocol for the contribution $\hat{a}_{x,y,+}^{\dagger} \hat{a}_{x,y,-}^{\nodagger}$ with untransformed density measurements on the remaining sites to access $f^{(R)}(\{\hat{n}\})$, the unnormalized denominator can then be sampled by computing
\begin{widetext}
\begin{equation}
    \vert \braket{\hat{\rho}^{(R)}_{\vec{j},{\vec{k}}}} \vert = \frac{1}{2}\sqrt{\langle \hat{j}_{x,y}(\nicefrac{\pi}{2}) f^{(R)}(\{\hat{n}\};\nicefrac{\pi}{2}) \rangle^2 + \langle\hat{j}_{x,y}(0) f^{(R)}(\{\hat{n}\};0) \rangle^2}.
\end{equation}
\end{widetext}

\paragraph{Evaluating the normalization (denominator):}
By construction, $\hat{U}_{\vec{j}}$ is a function solely of the occupation number operators $\hat{n}_{m}$ of the surrounding particles.
The expectation values in the denominator,$\braket{\hat{U}^{\dagger}_{\vec{j}}\hat{U}^{\nodagger}_{\vec{j}}}$ and $\braket{\hat{U}^{\nodagger}_{\vec{k}}\hat{U}^{\dagger}_{\vec{k}}}$, are therefore strictly diagonal in the Fock basis and do not require the local tunneling pulses used for the numerator. 
Instead, they can be directly computed from standard, independent quantum gas microscope snapshots of the atomic density distribution.
By evaluating the classical function $|U_{\vec{j}}(\{\hat{n}\})|^2$ over a sufficiently large ensemble of snapshots, the normalization factors $\braket{\hat{U}^{\dagger}_{\vec{j}}\hat{U}^{\nodagger}_{\vec{j}}}$ and $\braket{\hat{U}^{\nodagger}_{\vec{k}}\hat{U}^{\dagger}_{\vec{k}}}$ are obtained.
In summary, by combining one set of interferometric snapshots (for the off-diagonal numerator) with standard density snapshots (for the diagonal normalization factors), the exact Read composite boson correlator can be systematically reconstructed in state-of-the-art cold atom quantum simulators.


\renewcommand{\theequation}{S\arabic{equation}}
\renewcommand{\thefigure}{S\arabic{figure}}
\renewcommand{\thetable}{S\arabic{table}}


\setcounter{equation}{0}
\setcounter{figure}{0}
\setcounter{table}{0}
\setcounter{section}{0}

\clearpage

\onecolumngrid
\section*{SUPPLEMENTAL MATERIAL:\\ Long-range order and composite boson condensation in lattice quantum Hall states}
\setcounter{page}{1}
\begin{center}
	F.~J.~Pauw,$^{1,2}$ 
	N.~Goldman,$^{3,4,5}$ and
	F.~A.~Palm,$^{3,4}$\\
	\textit{$^{1}$ Department of Physics and Arnold Sommerfeld Center for Theoretical Physics (ASC), Ludwig-Maximilians-Universit\"at M\"unchen, Theresienstr. 37, D-80333 M\"unchen, Germany}\\
	\textit{$^{2}$ Munich Center for Quantum Science and Technology (MCQST), Schellingstr. 4, D-80799 M\"unchen, Germany}\\
	\textit{$^{3}$ CENOLI, Universit\'e Libre de Bruxelles, CP 231, Campus Plaine, B-1050 Brussels, BelgiumCenter for Nonlinear Phenomena and Complex Systems, Universit\'e Libre de Bruxelles, CP 231, Campus Plaine, B-1050 Brussels, Belgium}\\
	\textit{$^{4}$ International Solvay Institutes, B-1050 Brussels, Belgium}\\
	\textit{$^{5}$ Laboratoire Kastler Brossel, Coll\`ege de France, CNRS, ENS-Universit\'e PSL, Sorbonne Universit\'e, 11 Place Marcelin Berthelot, F-75005 Paris, France}
\end{center}

\twocolumngrid

\section{Numerical Details}
All numerical results in this Letter are obtained from density matrix renormalization group (DMRG)~\cite{White1992,Schollwock2005,Schollwock2011} simulations.
Specifically, we employ the single-site variant augmented with subspace expansion~\cite{Hubig2015} implemented in the \textsc{SyTen}~\cite{Syten} toolkit, explicitly exploiting the U$(1)_{n}$ symmetry associated with particle-number conservation.
\subsection{Convergence}
Convergence of the variational ground state across all parameter regimes was rigorously tracked via the truncation error and---for select configurations---the variance of the energy, $\mathrm{Var}(\hat{\mathcal{H}})$.
We performed sweeps until the discarded weight fell below a threshold of $10^{-8}$, thus requiring a maximum bond dimension of up to $\chi = 2048$.
Similarly, to rigorously assure convergence for the considered discarded weight, we monitored $\mathrm{Var}(\hat{\mathcal{H}})$ for the most challenging systems ($L=10$ for $n_{\rm max}=3$ soft-core bosons at $U/t=1.7 \approx U_{\rm c}/t$, and $L=14$ for both fermions and hard-core bosons).
Furthermore, as we are primarily interested in the long-range correlations of the matrix product state (MPS) ground state, we simultaneously monitored the convergence of the correlation length $\xi$.
The convergence behavior in this metric as a function of the bond dimension is shown in~\cref{figS:Conv}.

\begin{figure}[b]
	\centering
	\includegraphics{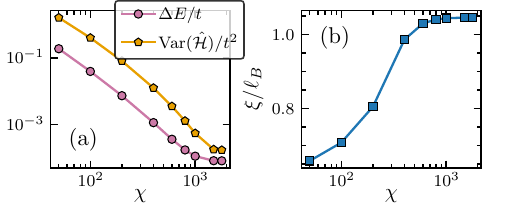}
	\caption{(a) Variance of the ground state energy and its deviation from the ground state energy extrapolated to infinite bond dimension as a function of bond dimension for fermions at $\nu=\nicefrac{1}{3}~(N=9)$ and $V/t=10$ for the largest considered system size $L=14$.
		(b) Correlation length in units of the magnetic length as a function of bond dimension.}
	\label{figS:Conv}
\end{figure}

\begin{figure*}[ht]
	\centering
	\includegraphics{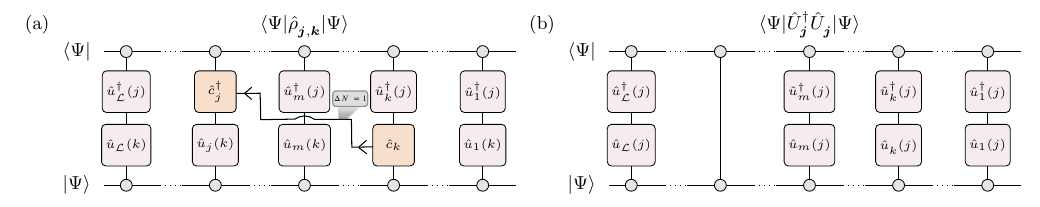}
	\caption{
		(a) Tensor network diagram illustrating the evaluation of the nonlocal string operators at the MPS-MPO level.
		The black arrow connecting the creation and annihilation operator accounts for the flow of the U$(1)$ quantum number.
		(b) Normalization of the non-unitary MPO by computing the norm of the nonlocal operator $\hat{U}^{(\dagger)}$.
	}
	\label{figS:MPO}
\end{figure*}
\subsection{MPO representation of nonlocal operators}
The direct evaluation of both the Read [\cref{eq:read_corr} of the main text] and the Girvin-MacDonald [\cref{eq:gm_corr} of the main text] composite boson correlation functions can be achieved by implementing the nonlocal string operator as a matrix product operator (MPO), much in the fashion of how Jordan-Wigner strings are encoded to promote hard-core bosonic tensor network states to fermionic ones.
Specifically, let \mbox{$\hat{b}_{ \vec{j}}=\hat{U}_{\vec{j}}\hat{a}_{\vec{j}}=\prod_{\vec{l}\neq \vec{j}}\hat{u}_{\vec{l}}(\vec{j})\hat{a}_{\vec{j}}$} be the full composite boson operator which acts on all lattice sites, where the phase string component $\hat{u}_{\vec{l}}(\vec{j})=\mathrm{e}^{i \vartheta_{\vec{l}}(\vec{j})\hat{o}_{\vec{l}}}$ acts on site $\vec{l} \neq \vec{j}$.
Here, $\vartheta$ is a function of the complex coordinate representation of the lattice points, and $\hat{o}$ is some local operator (e.g., $\hat{n}$, as considered throughout this Letter).
To compute the two-point correlation functions considered here, we evaluate the product:
\begin{widetext}
	\begin{equation}
		\label{eqS:GOperator}
		\hat{\rho}_{\vec{j},\vec{k}}=\hat{b}^{\dagger}_{\vec{j}}\hat{b}^{\nodagger}_{\vec{k}}=\hat{a}^{\dagger}_{\vec{j}}\prod_{\vec{l}\neq \vec{j}}\hat{u}^{\dagger}_{\vec{l}}(\vec{j})\prod_{\vec{l}\neq \vec{k}}\hat{u}^{\nodagger}_{\vec{l}}(\vec{k})\hat{a}^{\nodagger}_{\vec{k}}=
		\left(\prod_{\vec{l}< \vec{j}} \hat{u}^{\dagger}_{\vec{l}}(\vec{j})\hat{u}^{\nodagger}_{\vec{l}} (\vec{k})\right)\hat{a}^{\dagger}_{\vec{j}}\hat{u}^{\nodagger}_{\vec{j}}(\vec{k})\left(\prod_{\vec{j} <\vec{l}< \vec{k}} \hat{u}^{\dagger}_{\vec{l}}(\vec{j})\hat{u}^{\nodagger}_{\vec{l}}(\vec{k})\right)  \hat{u}^{\dagger}_{\vec{k}}(\vec{j})\hat{a}^{\nodagger}_{\vec{k}}\left(\prod_{\vec{k} <\vec{l}} \hat{u}^{\dagger}_{\vec{l}}(\vec{j})\hat{u}^{\nodagger}_{\vec{l}}(\vec{k})\right),
	\end{equation}
\end{widetext}
where we assume $j < k$, having established a bijective mapping $j = \mathcal{I}(\vec{j})$ from the two-dimensional lattice coordinates to the one-dimensional MPS path \mbox{$j, k \in \{1, \dots, \mathcal{L}\}$}.
\cref{eqS:GOperator} is a product of strictly local operators, which allows for an efficient representation in terms of single-site tensors.
The MPO construction proceeds by assigning a single local operator matrix, $\hat{\rho}^{[m]}$, to each site $m$ along the MPS path (where $\vec{l} = \mathcal{I}^{-1}(m)$):
\begin{enumerate}
	\item At all bulk sites $m \notin \{j, k\}$, the tensor applies the combined phase and amplitude modulation: \\ $\hat{\rho}^{[m]} = \hat{u}^{\dagger}_{\vec{l}}(\vec{j})\hat{u}^{\nodagger}_{\vec{l}}(\vec{k}) =\mathrm{e}^{i (\vartheta_{\vec{l}}(\vec{k}) - \vartheta^*_{\vec{l}}(\vec{j})) \hat{o}_{\vec{l}}}$.
	\item At the first anchor site $j$, the tensor absorbs the creation operator alongside the remaining phase string: $\hat{\rho}^{[j]} = \hat{a}^{\dagger}_{\vec{j}}\hat{u}^{\nodagger}_{\vec{j}}(\vec{k})$.
	\item At the second anchor site $k$, the tensor absorbs the annihilation operator: $\hat{\rho}^{[k]} = \hat{u}^{\dagger}_{\vec{k}}(\vec{j})\hat{a}^{\nodagger}_{\vec{k}}$.
\end{enumerate}
If $\hat{o}=\hat{n}$, evaluating the exponential in the occupation number basis yields a purely diagonal matrix.
Alternatively, because the chosen local operator $\hat{n}$ possesses eigenvalues bounded by a maximum occupancy $n_{\max}$, the matrix exponentials can be evaluated exactly via a discrete polynomial expansion in $\hat{n}$. 
Finally, the full two-point correlator is obtained by calculating the expectation value $\langle \Psi | \hat{\rho}^{[1]} \otimes \dots \otimes \hat{\rho}^{[\mathcal{L}]} | \Psi \rangle$, which is contracted efficiently by sweeping through the MPS in $\mathcal{O}(\mathcal{L} \chi^3)$ operations, where $\chi$ is the MPS bond dimension and $\mathcal{L}$ is the total number of lattice sites.
This process is illustrated in \cref{figS:MPO}.

Importantly, while the GM operator is unitary, evaluating the non-unitary Read MPO requires the MPS expectation value to be explicitly normalized with respect to the action of the nonlocal constituents of the operator:
\begin{equation}
	\frac{\braket{\Psi| \hat{\rho}^{\nodagger}_{\vec{j},\vec{k}} |\Psi}}{\sqrt{\braket{\Psi|\hat{U}^{\dagger}_{\vec{j}}\hat{U}^{\nodagger}_{\vec{j}}|\Psi} \braket{\Psi|\hat{U}^{\nodagger}_{\vec{k}}\hat{U}^{\dagger}_{\vec{k}}|\Psi}}}.
\end{equation}

To keep the numerical costs manageable, we employed a bond dimension truncation of the MPS ground state prior to evaluating the full system two-point correlation matrix $\rho^{(\cdot)}$.
While optimizing the ground state energy typically requires large bond dimensions, the nonlocal topological features and long-range correlations are highly robust and converge at significantly lower effective bond dimensions, $\chi_{\rm eff}$.
Therefore, the maximum truncation cut-off is chosen such that the relative error in the Frobenius norm, \mbox{$\Delta_F(\chi)= \| \rho^{(\cdot)}(\chi) - \rho^{(\cdot)}(\chi_{\rm eff}) \|_F / \| \rho^{(\cdot)}(\chi_{\rm eff}) \|_F$}, falls below a strict tolerance threshold of $10^{-3}$, ensuring that the extracted condensate fractions remain physical under the compression.
This truncation process is illustrated in \cref{figS:ReadConv}.
For example, in the ground state of a $10 \times 10$ square lattice of soft-core bosons ($n_{\rm max}=3$) at half-filling $\nu=1/2$ and intermediate on-site repulsion $U/t$, the energy converges at a bond dimension $\chi_{\rm conv}\approx1000$, while the topological features of the ground state are already fully converged at $\chi_{\rm eff}\approx 300$.

\begin{figure}[t]
	\centering
	\includegraphics{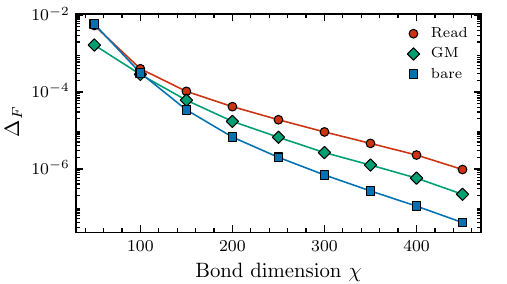}
	\caption{Error estimation of the full-system two-point correlation matrix $\rho^{(\cdot)}$ (bare, GM, and Read) as a function of bond dimension in terms of the Frobenius norm distance with respect to a reference bond dimension $\chi_{\rm conv}\approx 1000$ for a $10\times 10$ lattice of soft-core bosons ($n_{\rm max}=3$) at $\nu=\nicefrac{1}{2}~(N=6)$.
	}
	\label{figS:ReadConv}
\end{figure}

\begin{figure*}
	\centering
	\includegraphics{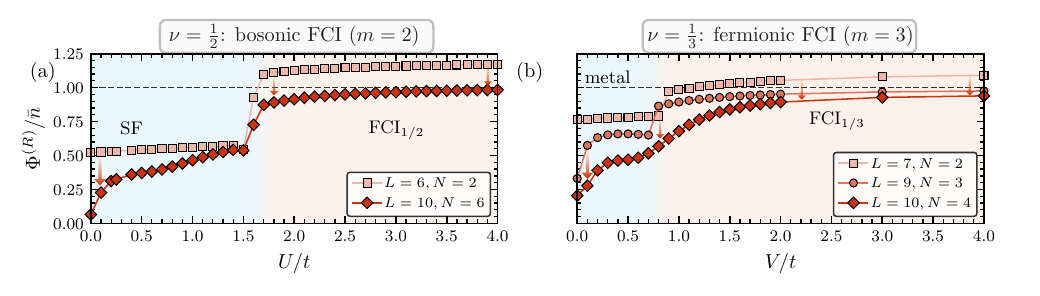}
	\caption{Normalized Read order parameter $\Phi^{(\rm R)}/\bar{n}$ as a function of interaction for varying system sizes.
		Panel (a) shows bosons with $n_{\rm max}=3$ at $\nu=\nicefrac{1}{2}$ where the on-site repulsion $U/t$ is tuned.
		Panel (b) shows fermions at $\nu=\nicefrac{1}{3}$ where the nearest-neighbor interaction $V/t$ is varied.
	}
	\label{figS:PT}
\end{figure*}

\section{Finite Size Analysis of the phase transitions}
While in the continuum the Read order parameter is expected to show a sharp transition from a zero to a non-zero valued $\Phi^{(R)}$, the finite-size nature of the lattice construction shows a broadening of this behavior.
To demonstrate that in the infinite system-size limit this sharp transition is recovered, we analyze the order parameter along the SF/metal to FCI phase transition for increasing system sizes.
In Fig.~\ref{figS:PT}(a), we plot the normalized Read order parameter, $\Phi^{(R)}/\overline{n}$, for a bosonic system at filling factor $\nu=1/2$ (corresponding to $m=2$) across the superfluid (SF) to FCI transition.
The phase transition is driven by the on-site interaction strength $U/t$, allowing for a maximum local occupancy of $n_{\max}=3$.
For small system sizes, such as $L=6$ ($N=2$), the transition is heavily affected by finite-size effects.
However, as the system size is systematically increased up to $L=10$ ($N=6$), the crossover visibly sharpens, confirming that the topological order parameter rapidly saturates toward its expected value in the deep FCI regime while vanishing in the trivial SF phase.
A similarly consistent finite-size scaling behavior is observed for the fermionic case at filling $\nu=1/3$ ($m=3$), as shown in Fig.~\ref{figS:PT}(b).
Here, the transition from a gapless metal to the fermionic FCI is driven by the nearest-neighbor repulsion $V/t$.
As the system size is scaled from $L=7$ ($N=2$) to $L=10$ ($N=4$), the normalized order parameter exhibits a progressively steeper slope near the critical interaction strength.
Together, these finite-size scaling results strongly support the validity of the Read composite boson operator as a reliable indicator of topological order, correctly recovering the expected sharp phase boundaries in the thermodynamic limit.

\section{Statistics of the Composite Operators}
We continue by showing that both the Read and Girvin-MacDonald (GM) composite operators, $\hat{b}^{(\dagger)}_{\cdot, \vec{j}}$, exhibit bosonic statistics on the lattice by explicitly evaluating their commutation relations---specifically  \mbox{$[\hat{b}_{\vec{j}}, \hat{b}_{\vec{k}}] = 0$} and $[\hat{b}^{\nodagger}_{\cdot,\vec{j}}, \hat{b}^\dagger_{\cdot, \vec{k}}] = 0$ for $\vec{j} \neq \vec{k}$. 
Both the GM and Read operators construct the composite particle by attaching a string of operators to the bare particle $\hat{a}_{\vec{j}}$:
\begin{equation}
	\hat{b}_{\cdot, \vec{j}} = \hat{U}_{\cdot, \vec{j}} \hat{a}_{\vec{j}}, \quad \text{where} \quad \hat{U}_{\cdot, \vec{j}} = \prod_{\vec{l} \neq \vec{j}} \hat{u}_{\cdot, \vec{l}}(\vec{j}).
\end{equation}
For both operators, the component acting on site $\vec{l}$ has the form:
\begin{equation}
	\hat{u}_{\cdot, \vec{l}}(\vec{j}) = \mathrm{A}_{\cdot,\vec{l}\vec{j}} \mathrm{e}^{i m \arg(w_{\vec{j}} - w_{\vec{l}}) \hat{n}_{\vec{l}}},
\end{equation}
where $m$ is the topological charge and $\mathrm{A}_{\cdot, \vec{l}\vec{j}} = \mathrm{A}_{\cdot, \vec{j}\vec{l}}$ is a purely real, symmetric amplitude factor. 
Because $\mathrm{A_{\cdot}}$ is symmetric and depends strictly on local densities, it commutes with all other density-dependent terms. 
We therefore focus entirely on the phase factor.
The product of two composite operators at different sites yields:
\begin{equation}
	\hat{b}_{\cdot, \vec{j}} \hat{b}_{\cdot,\vec{k}} = \hat{U}_{\cdot, \vec{j}} \hat{a}_{\vec{j}} \hat{U}_{\cdot, \vec{k}} \hat{a}_{\vec{k}}.
\end{equation}
To move $\hat{a}_{\vec{j}}$ past $\hat{U}_{\cdot, \vec{k}}$, we must commute it through the phase string.
The only term in $\hat{U}_{\cdot, \vec{k}}$ that does not trivially commute with $\hat{a}_{\vec{j}}$ is the phase acting on site $\vec{j}$, which is $\mathrm{e}^{i m \arg(w_{\vec{k}} - w_{\vec{j}}) \hat{n}_{\vec{j}}}$. 
Using the fundamental algebraic identity for creation and annihilation operators, $\hat{a}_{\vec{j}} f(\hat{n}_{\vec{j}}) = f(\hat{n}_{\vec{j}} + 1) \hat{a}_{\vec{j}}$, we find:
\begin{equation}
	\hat{a}_{\vec{j}} \, \mathrm{e}^{i m \arg(w_{\vec{k}} - w_{\vec{j}}) \hat{n}_{\vec{j}}} = \mathrm{e}^{i m \arg(w_{\vec{k}} - w_{\vec{j}})} \mathrm{e}^{i m \arg(w_{\vec{k}} - w_{\vec{j}}) \hat{n}_{\vec{j}}} \hat{a}_{\vec{j}}.
\end{equation}
Therefore, commuting $\hat{a}_{\vec{j}}$ past $\hat{U}_{\cdot, \vec{k}}$ extracts a scalar phase, yielding:
\begin{equation}
	\hat{b}_{\cdot,\vec{j}} \hat{b}_{\cdot, \vec{k}} = \mathrm{e}^{i m \arg(w_{\vec{k}} - w_{\vec{j}})} \hat{U}_{\cdot, \vec{j}} \hat{U}_{\cdot, \vec{k}} \hat{a}_{\vec{j}} \hat{a}_{\vec{k}}.
\end{equation}
By symmetry, reversing the order of the operators yields:
\begin{equation}
	\hat{b}_{\cdot, \vec{k}} \hat{b}_{\cdot, \vec{j}} = \mathrm{e}^{i m \arg(w_{\vec{j}} - w_{\vec{k}})} \hat{U}_{\cdot, \vec{k}} \hat{U}_{\cdot, \vec{j}} \hat{a}_{\vec{k}} \hat{a}_{\vec{j}}.
\end{equation}
Because $\hat{U}_{\cdot, \vec{j}}$ and $\hat{U}_{\cdot, \vec{k}}$ depend only on density operators, they inherently commute. 
This leaves us with:
\begin{align}
	\mathrm{e}^{i m \arg(w_{\vec{k}} - w_{\vec{j}})} \hat{a}_{\vec{j}} \hat{a}_{\vec{k}} & = \mathrm{e}^{i m \arg(w_{\vec{j}} - w_{\vec{k}})} \hat{a}_{\vec{k}} \hat{a}_{\vec{j}} \nonumber \\
	\mathrm{e}^{\pm i m \pi} \mathrm{e}^{i m \arg(w_{\vec{j}} - w_{\vec{k}})} \hat{a}_{\vec{j}} \hat{a}_{\vec{k}} &= \mathrm{e}^{i m \arg(w_{\vec{j}} - w_{\vec{k}})} \hat{a}_{\vec{k}} \hat{a}_{\vec{j}} \nonumber \\
	\implies \mathrm{e}^{i m \pi} \hat{a}_{\vec{j}} \hat{a}_{\vec{k}} &= \hat{a}_{\vec{k}} \hat{a}_{\vec{j}},
	\label{eqS:StatExchange}
\end{align}
where we used $\arg(w_{\vec{j}} - w_{\vec{k}})= \arg(w_{\vec{k}} - w_{\vec{j}})\pm \pi$.
Equation~\eqref{eqS:StatExchange} perfectly demonstrates the statistical transmutation.
The final commutation behavior depends strictly on the intrinsic statistics of the bare particles ($\hat{a}_{\vec{j}}$) and the topological charge ($m$) independent of attaching quasiholes (Read) or flux quanta (GM).
For the bosonic fractional Chern insulator state ($m=2$), the underlying particles are bosons, which commute via $\hat{a}_{\vec{j}} \hat{a}_{\vec{k}} = (+1) \hat{a}_{\vec{k}} \hat{a}_{\vec{j}}$ and hence, equation~\eqref{eqS:StatExchange} becomes $\mathrm{e}^{i 2 \pi} (+1) = 1$.
In contrast, for the fermionic fractional Chern insulator ($m=3$), the underlying particles are fermions, which anticommute via $\hat{a}_{\vec{j}} \hat{a}_{\vec{k}} = (-1) \hat{a}_{\vec{k}} \hat{a}_{\vec{j}}$ and consequently, equation~\eqref{eqS:StatExchange} evaluates to $\mathrm{e}^{i 3 \pi}(-1) =  1$.
Attaching an odd number of fluxes to a fermion completely cancels the Pauli minus sign, successfully transmuting the composite object into a boson.
Completely analogously, it follows that $[\hat{b}^{\nodagger}_{\cdot, \vec{j}}, \hat{b}_{\cdot, \vec{k}}^\dagger] = 0$ by utilizing $\hat{a}_{\vec{k}}^\dagger f(\hat{n}_{\vec{k}}) = f(\hat{n}_{\vec{k}} - 1) \hat{a}_{\vec{k}}^\dagger$, which simply conjugates the phase factors but preserves the fundamental $\mathrm{e}^{i m \pi}$ exchange symmetry.

\end{document}